\DeclareMathAlphabet{\mathscr}{OT1}{pzc}{m}{it}

\documentclass[prb,footinbib,twocolumn,notitlepage]{revtex4-1}
\usepackage{graphicx}
\usepackage{epsfig}
\usepackage{dcolumn}
\usepackage{bm}
\usepackage{amsmath}
\usepackage{amsbsy}
\usepackage{amssymb}
\usepackage[usenames]{color}
\usepackage{mathrsfs}
\usepackage{verbatim}
\usepackage{hyperref}

\newcommand{\bra}[1]{\ensuremath{\langle{#1}|\,}}
\newcommand{\ket}[1]{\ensuremath{\,|{#1}\rangle}}

\begin{document}


\title{Light-hole transitions in quantum dots: realizing full control by highly focused optical-vortex beams}

\author{G. F. Quinteiro$^{1,2}$}
\email{gquinteiro@df.uba.ar}
\author{T. Kuhn$^2$}

\affiliation{
$^1$Departamento de F\'isica and IFIBA, FCEN,
Universidad de Buenos Aires, Ciudad Universitaria, Pabell\'on I,
1428 Ciudad de Buenos Aires, Argentina \\
$^2$ Universit\"at M\"unster,
Wilhelm-Klemm-Str. 10, 48149 M\"unster, Germany
}

\date{\today}

\begin{abstract}
An optical-vortex is an inhomogeneous light beam having a phase singularity at its axis, where the intensity of the electric and/or magnetic field may vanish.  
Already well studied are the paraxial beams, which are known to carry well defined values of spin (polarization $\sigma$) and orbital angular momenta; the orbital angular momentum per photon is given by the topological charge $\ell$ times the Planck constant. 
Here we study the light-hole--to--conduction band transitions in a semiconductor quantum dot induced by a highly-focused beam originating from a $\ell=1$ paraxial optical vortex. 
We find that at normal incidence the pulse will produce two distinct types of electron--hole pairs, depending on the relative signs of $\sigma$ and $\ell$.
When sign($\sigma$)$=$sign($\ell$), the pulse will create electron--hole pairs with band+spin and envelope angular momenta both equal to one.
In contrast, for sign($\sigma$)$\neq$sign($\ell$), the electron-hole pairs will have neither band+spin nor envelope angular momenta.
A tightly-focused optical-vortex beam thus makes possible the creation of pairs that cannot be produced with plane waves at normal incidence.
With the addition of co-propagating plane waves or switching techniques to change the charge $\ell$ both the band+spin and the envelope angular momenta of the pair wave-function can be precisely controlled.
We discuss possible applications in the field of spintronics that open up.
\end{abstract}

\pacs{42.50.Tx, 78.67.-n}

\maketitle


\section{Introduction}
\label{Sec:Intro}

Quantum dots (QDs) are artificial systems that confine particles in all directions, yielding a discrete spectrum of energy levels. A variety of QDs have been fabricated and studied, such as laterally confined QDs in a two-dimensional electron gas, colloidal semiconductor nanocrystals, and semiconductor self-assembled QDs to name only few. 
Due to their semiconductor nature, self-assembled QDs have electron energy levels grouped into valence (with heavy- and light-hole subbands) and conduction bands. In most cases, the upper-most energy levels in the valence band are of heavy-hole (HH) type.
The energy difference between bands makes it possible to excite electrons from the valence to the conduction band by optical or near-optical fields. For this reason, applications to spintronics typically focus on the fast control of the spin state of carriers or nearby impurities by short-pulse excitation of QD states. The studies mostly consider excitation from HH states, since decaying is reduced by the lack of lower energy levels.
The recent literature shows an increasing interest in light-hole (LH) state excitation and manipulation\cite{feng2004spin, reiter2011coherent, pazy2003spin}. A strong advantage of LHs over HHs is that, with their spin projections $\pm 1/2$, spin-flips with e.\ g.\ an impurity spin can occur\cite{reiter2011coherent}. However, the fast decay channel leading to the lower HH levels makes it difficult to use LHs in proposals designed for the former. Recently Huo et al.\cite{huo2013light} demonstrated energy inversion between hole subbands in stressed GaAs QDs. Their successful experiment widens the applicability of LHs to spintronics. 

Strongly varying electromagnetic fields attract the attention of researchers in diverse areas such as optics, atomic and molecular physics, and information technology\cite{Andrews2008Structure}. A prominent example is that of optical-vortex fields, characterized by a phase singularity where the intensity of the electric and/or magnetic field may vanish. Instances of beams with optical vortex are the radially/azimuthally-polarized beams and twisted-light (TL) Bessel and Laguerre-Gaussian beams. It has been shown that a paraxial optical vortex carries orbital angular momentum (AM) equal to $\hbar \ell$, where $\ell$ is the topological charge. However, when such a beam is tightly focused, orbital and spin AM are not well separated\cite{barnett1994orbital} and the beam becomes a superposition of modes with different topological charge\cite{bliokh2011spin}.
In solid state systems, it has been shown that paraxial beams of TL bring about new effects such as circular photon-drag effect in bulk\cite{quinteiro_theory_2009} and nanostructures\cite{quinteiro2009ele2, watzel2012pho}, new optical transitions in quantum dots\cite{quinteiro_79_155450}, excitation of new exciton states\cite{ueno2009coh, quinteiro2010bel, shigematsu2013orb}, and excitation of intersubband transitions at normal incidence in quantum wells\cite{sbierski2013twi}. 
A number of authors have called attention to a special set of non-paraxial optical-vortex fields. They have shown that the values $\ell=1,2$ (to be understood as the value of the corresponding paraxial original beam) present specially interesting features\cite{monteiro2009ang, klimov2012mapping, iketaki2007inv, quinteiro2014formulation}. For instance, close to the phase singularity and for circular polarization $\sigma=-1$ (polarization vector $\hat{x}-i\hat{y}$) and $\ell=2$ the magnetic interaction overcomes the electric interaction. Or for the set $\{\ell=1,\sigma=-1\}$ a non-vanishing longitudinal component of the electromagnetic field exists even at the phase singularity.

In this article we study the excitation of LHs in semiconductor QDs by pulses of highly-focused beams of optical vortices with emphasis on the value $\ell=1$. We show that by only changing the polarization state of the light, electron--hole (eh) pair states with and without band+spin and envelope AM can be addressed at normal incidence. Moreover, in conjunction with a co-propagating field with topological charge $\ell = 0$, all states formed out of envelope $s$ and $p$ shells can be controlled. Compared to proposals using two orthogonally propagating pulses of plane waves, our normal-incidence configuration may significantly improve the experimental realization of LH states control.

The article is organized as follows. Sections \ref{Sec:System} and \ref{Sec:ell1} introduce the physical system and electromagnetic fields, respectively. The optical transitions are discussed in Sect.\ \ref{Sec:trans}. Possible applications to spintronics are given in Sect.\ \ref{Sec:spintronics}. The article ends with the conclusions in Sect.\ \ref{Sec:conc}.

\section{Quantum dot}
\label{Sec:System}

Semiconductor self-assembled QDs can be fabricated to confine electrons and holes. For small QDs the energy spectrum is a collection of discrete states grouped in two valence (LH and HH) bands and one conduction band. The valence bands are characterized by a total angular momentum $j=3/2$. In the HH band the $z$ components are given by $m_j=\pm 3/2$ while in the LH bands they are $m_j=\pm 1/2$. Here we neglect possible band mixing effects caused by strain distributions of the Luttinger Hamiltonian. In the conduction band $j=s=1/2$ and $m_s=\pm1/2$. Conduction and valence bands are separated by a band-gap energy $E_g$ in the few eV range; thus, interband transitions may be produced by optical or near-optical excitation.

In general, the upper-most energy levels in the valence band are of HH type; however, an inversion between LH and HH bands is possible by applying stress to the sample\cite{huo2013light}. The inversion of energy levels is advantageous in proposals using LH excitations, since it slows the decay of the states.

QDs can be fabricated out of different materials. Typical ones are InAs/GaAs or CdTe/ZnTe QDs, but others may prove to be more suitable in certain situations. In particular, QDs made of GaN/AlN\cite{kako2004exciton, andreev2000theory} have a large $E_g$ and therefore are excited at shorter wavelength. We will later discuss the possible advantage of these QDs.

We assume that the growth direction of the QD coincides with the $z$-axis, the QD has the shape of a disk with radius $r_0\simeq10$nm and height $z_0\simeq 0.1r_0$ and is centered at $r=0$. 
The complete wave function for carries in the valence and conduction bands are
\begin{eqnarray}
\label{Eq:complete_states}
    \psi(\mathbf r)
&=&
    \phi(r,\varphi) {\cal Z}(z) u_i(\mathbf r)
\,,
\end{eqnarray}
where the envelope functions are ${\cal Z}(z)$, $\phi_{isn}(r,\varphi) = {\cal R}_{isn}(r)(2\pi)^{-1/2}e^{i n \varphi}$, with $n$ being the angular quantum number. ${\cal R}_{isn}(r/\ell_c)$ is the radial component of the wave-function with confinement length $\ell_c$ and radial quantum number $s$. The detailed functional form of ${\cal R}_{isn}(r/\ell_c)$ depends on the confinement potential, which could be e.\ g.\ square or parabolic (for the latter see Ref.\ \onlinecite{quinteiro_79_155450} noticing a change in sign).
The Bloch-periodic parts of electron states in the LH band are\cite{Bastard1988Wave}
\begin{eqnarray}
\label{Eq:LH_states}
    \ket{3/2, +1/2}
&=&
    -\frac{1}{\sqrt{6}}
        \left[
            {
				{ 
				   	\left( \ket{p_x} + i \ket{p_y} \right) 
				}
				\downarrow
			}
            - 2\ket{p_z} \uparrow
        \right]
\nonumber \\
    \ket{3/2, -1/2}
&=&
    \frac{1}{\sqrt{6}}
        \left[
            \left( \ket{p_x} - i \ket{p_y} \right) \uparrow
            + 2\ket{p_z} \downarrow
        \right]
\,,
\end{eqnarray}
alternatively we will refer to these states as $u_\pm (\mathbf r)$. The $s$-like conduction-band states are written $u_s(\mathbf r)= \langle \mathbf r \ket{s} \xi$, where $\xi$ is the spin.

In what follows, we will distinguish among different contributions to the angular momentum of carriers. For the sake of clarity we explicitly state them. Carrier have the {\it band} AM of their energy band: in the case of the $p$-like valence band this is equal to $1$, while for the $s$-type conduction band this is equal to $0$. In addition, electrons have {\it spin} AM equal to $1/2$. These two comprise the Bloch periodic-part or band+spin AM [Eq.\ (\ref{Eq:LH_states})]. Finally, there is an {\it envelope} AM $\hbar n$ corresponding to the envelope part of the wave-function.

\section{Optical-vortex field}
\label{Sec:ell1}

To be specific in what follows we use TL Bessel beams\cite{quinteiro_theory_2009}, which are one class of optical-vortex fields. However, much of our findings can be applied to other optical-vortex beams, as we explicitly discuss at the end of this section.

We study a monochromatic field of TL with wave-length $\lambda \simeq 500$nm and circular polarization $\sigma=\pm 1$ propagating along the QD growth axis (i.\ e.\ impinging the sample at normal incidence). We use a non-paraxial expression for the vector potential\cite{quinteiro_theory_2009}
\begin{eqnarray}
\label{Eq:A_Bessel}
	A_x(\mathbf r, t)
&=&
	F_{\ell}(r) \cos[(\omega t-q_z z) -	\ell \varphi]
\nonumber \\
	A_y(\mathbf r, t)
&=&
	\sigma F_{\ell}(r) \sin[(\omega t-q_z z) - \ell \varphi]
\nonumber \\
	A_z(\mathbf r, t)
&=&
	-\sigma \frac{q_r}{q_z} F_{\ell+\sigma}(r)
	\sin[(\omega t-q_z z) - (\ell + \sigma)\varphi]
\,.\,\,
\end{eqnarray}
With $F_{\ell}(r)$ being a Bessel function, the vector potential satisfies the Coulomb gauge condition and the vectorial Helmholtz equation. Close to the phase singularity one can approximate $F_{\ell}(r)\simeq A_0 (q_r r)^{\ell}/\ell!$. In the paraxial limit ($q_r/q_z \ll 1$) $A_z$ is negligible, then $\ell$ is the topological charge and the beam carries a single value of orbital AM. However, for a non-paraxial beam $\sigma$ and $\ell$ mix, and while $\{A_x,A_y\}$ retain their a value $\ell$ for the topological charge, $A_z$ has a charge equal to $\ell+\sigma$. 
In what follows we will assume $\ell=1$, but one can naturally choose $\ell=-1$, since the important distinction is that of the relative signs of $\sigma$ and $\ell$. Furthermore the beam is assumed to be highly focused such that $q_r/q_z\simeq 1$, where $\sqrt{q_r^2+q_z^2}=2\pi/\lambda$, and $q_r\simeq 10^{-2}$nm$^{-1}$ is a measure of the beam waist.

Elsewhere\cite{quinteiro2014formulation} we have shown that a convenient mathematical representation of TL close to the phase singularity (located at $r=0$) is obtained by the so-called TL-gauge. This gauge transformation casts the Hamiltonian into an all-electric dipolar-like Hamiltonian when sign$(\ell)\!=$sign$(\sigma)$. For light with sign$(\ell)\!\neq$sign$(\sigma)$, magnetic terms appear that should not be a priori neglected. Magnetic terms indeed overcome electric terms, for particular values of $\ell$. 
However, for $\{\ell=1,\sigma=-1\}$ and the parameters chosen for the QD and beam all magnetic contributions are negligible (see the Appendix) and we can indeed use the TL-gauge Hamiltonian. The direct consequence is that we can restrict our analysis to just electric interactions for both polarizations.

For light with polarization $\sigma=-1$ [using $E=-\partial_t A$ and Eq.\ (\ref{Eq:A_Bessel}) with $F_{\ell}(r)\simeq A_0 (q_r r)^{\ell}/\ell!$]
\begin{eqnarray}
\label{Eq:AM0_Efield_z}
	E_x
&=&
	\frac{E_0}{2} (q_r r) \sin(\omega t - q_z z - \varphi) \,,
\nonumber \\
	E_y
&=&
	\frac{E_0}{2} (q_r r) \cos(\omega t - q_z z - \varphi) \,,
\nonumber \\
	E_z
&=&
	- E_0 \frac{q_r}{q_z} \cos(\omega t - q_z z)
\,,
\end{eqnarray}
with $E_0=\omega A_0$. The dominant term is the $z$ component, because the others are a factor $(q_r r) < 0.1$ smaller. The Hamiltonian reads
\begin{eqnarray}
\label{Eq:H_AM0}
	H
&=&
	\frac{1}{2m} \mathbf p^2 + V(\mathbf r)
    - d_z E_z(\mathbf r,t)
\,,
\end{eqnarray}
where the dipole moment is $d_z=q \, z$, and we assume that the canonical momentum $\mathbf p$ is equal to the mechanical momentum.

For light with polarization $\sigma=+1$ 
\begin{eqnarray}
\label{Eq:Fields_OAM_cart}
    E_x(\mathbf r; t)
&=&
    \frac{E_0}{2} (q_r r) \sin(\omega t - q_z z -\varphi) \,,
\nonumber \\
    E_y(\mathbf r; t)
&=&
    -\frac{E_0}{2} (q_r r) \cos(\omega t - q_z z -\varphi) \,,
\nonumber \\
    E_z(\mathbf r; t)
&=&
    \frac{E_0}{8} (q_r r)^2 \cos(\omega t - q_z z - 2 \varphi) \,,
\end{eqnarray}
the in-plane components of the electric field dominate the interaction, with the Hamiltonian
\begin{eqnarray}
\label{Eq:Ham_OAM}
	H
&=&
    \frac{\mathbf p^2}{2m} +V(\mathbf r) 
	-\frac{1}{2} \mathbf d_\perp \cdot \mathbf E(\mathbf r; t)
\,,
\end{eqnarray}
where $\mathbf d_\perp=q (x \hat{x}+y \hat{y})$. The extra $1/2$ factor is a result of the vanishing of the field at the phase singularity\cite{quinteiro2014formulation}.

An interaction Hamiltonian such as \mbox{$H_I = - d_z E_z(\mathbf r; t)$} contains operators acting on the envelope and the periodic part of the wave-function.
This is clearly seen by writing $\mathbf r = \mathbf R + \mathbf r_c$, where $\mathbf r_c$ designates points within a crystal unit-cell and $\mathbf R$ designates different unit cells in the whole crystal. The change of variable leads to $H_I = - q (\mathbf R + \mathbf r_c) \cdot \mathbf E(\mathbf R + \mathbf r_c)$.
Since the electric field varies little within the unit cell we can disregard $\mathbf r_c$ in the argument of the electric field.
The same cannot be done for the dipole moment and indeed each term gives rise to different processes. 
{\it Inter}-band transitions are described by $H_I^{inter} = - q \mathbf r_c  \cdot \mathbf E(\mathbf R)$, while {\it intra}-band transitions by $H_I^{intra} = - q \mathbf R  \cdot \mathbf E(\mathbf R)$. 
We study in this article interband transitions. 

To conclude the section, we briefly comment on the similar behavior exhibited by other optical-vortex fields. 
The focusing of a paraxial optical vortex by a lens has been shown to produce a conversion between orbital and spin AM\cite{bliokh2011spin},  as it was commented after Eq.\ (\ref{Eq:A_Bessel}). Other authors have studied the specific case of focusing of an incoming Laguerre-Gaussian modes of TL with $\ell=1$ in experiments\cite{iketaki_investigation_2007} and theory\cite{monteiro2009ang, klimov2012mapping}. These works are in agreement with our non-paraxial electric field expressions Eq.\ (\ref{Eq:AM0_Efield_z}) and Eq.\ (\ref{Eq:Fields_OAM_cart}). In particular, they found that for $\sigma=-1$ only the $z$ component of the electric field survives at the phase singularity, while for $\sigma=1$ all components vanish. Calculations also show that the $r$ dependence of the field components is as in Eq.\ (\ref{Eq:AM0_Efield_z}) for both polarization states of the light and small $q_r r$.
In addition, radially/azimuthally polarized beams, which are superpositions of $\ell=\pm 1$\cite{ornigotti2013radially}, exhibit similar features. For example, for focused radially-polarized beams there is a non-vanishing $z$ component of the electric field at the phase singularity\cite{dorn2003sharper, youngworth2000focusing}.
Besides, Biss et al.\cite{biss2001cylindrical} have shown that radially/azimuthally polarized beams retain their spatial profile when focused through a dielectric interface, such as those of semiconductor heterostructures.  
This implies that the experimental control of the light-hole in a QD can be realize by focusing either Bessel or Laguerre-Gaussian TL modes. Also, focused radially-polarized beams can be used for the partial control of light-holes, as proposed with $\{\ell=1, \sigma=-1\}$ TL below. 
%

\section{Light-hole transition induced by an optical vortex}
\label{Sec:trans}

Unlike HHs, LHs present an admixture of spin and band AM, as seen in Eq.\ (\ref{Eq:LH_states}), that gives rise to richer optical transitions. The transitions create eh pairs. For strong confinement potential in QDs\cite{dzyubenko1993magnetoexcitons} the Coulomb interaction, that accounts for exciton complexes, can be treated perturbatively\cite{wojs1995negatively}. Since we only consider small QDs, we describe optical transitions without the Coulomb interaction--new features related to the interaction of TL and excitons in bulk were studied in Ref.\ \onlinecite{quinteiro2010bel, ueno2009coh, shigematsu2013orb}. 

\begin{figure}[h]
  \centerline{\includegraphics[scale=.6]{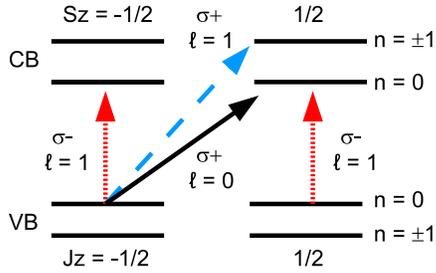}}
  \caption{Transitions induced by light having $\ell=0,1$ on an electron initially in the $\ket{3/2,-1/2}$ LH band. The $z$ component of light with $\{\ell=1, \sigma=-1\}$ produces an e-LH pair with AM$=$0 [Dotted (red)]. The in-plane component of light with $\sigma=+1$ produces an e-LH pair with AM$=$1 whose envelope AM is: $n=0$ for $\ell=0$ [solid(black)], and $n=1$ for $\ell=1$ [dashed(blue)].}
  \label{fig:Trans}
\end{figure}

When an in-plane electromagnetic field impinges on the QD, an eh pair is created, due to the first term in the decomposition of $\ket{3/2, \pm 1/2}$ in Eq.\ (\ref{Eq:LH_states}). The pair thus produced has an AM projection equal to $\pm 1$, due to the combination of conduction-band electron and LH band+spin AM. Moreover, if the light carries no orbital AM ($\ell=0$), the transition is the one shown in Fig.\ \ref{fig:Trans} by solid (black) line. In contrast, twisted light can transfer its orbital AM to the pair, by promoting an electron from an $s$-like envelope state in the valence band to a $p$-like envelope state in the conduction band as exemplified by the dashed (blue) line with $\ell=1$. 

The second term in the decomposition of hole states in Eq.\ (\ref{Eq:LH_states}) can only be addressed by a longitudinal $z$-component (parallel to the QD's growth axis) of the electric field. The resulting state is an eh pair with $\text{AM}=0$, see the dotted (red) lines in Fig.\ \ref{fig:Trans}.
Pairs with $\text{AM}=0$ can be excited by plane-wave light propagating along the $x-y$ plane\cite{reiter2011coherent}, although the corresponding experimental set-up is usually cumbersome or even not possible.

In the next sections we will show that twisted light can produce both types of eh pairs. 
This is done by calculating the transition matrix elements, which are essential to analyze excitation processes using for example the Fermi's Golden Rule.

\subsection{Electron--light-hole pairs with vanishing band+spin angular momentum}
\label{Sec:AM0}

The excitation of eh pair with $\text{AM}=0$ can be accomplished by twisted light having $\{\ell=1,\sigma=-1\}$, thanks to its strong $z$-component electric field [see Eq.\ (\ref{Eq:AM0_Efield_z})]. 
We evaluate the matrix elements for the transition between LH and conduction bands. Under the assumption of nearly resonance excitation the Rotating Wave Approximation is used and the matrix element for the absorption of a photon, with $H_I = -q z_c E_z(\mathbf Z; t)$ is 
\begin{eqnarray}
\label{Eq:AM0_ME}
    \bra{\psi_{c}} H_I \ket{\psi_{LH}}
&\simeq&
    q \frac{E_0}{2} \frac{q_r}{q_z} e^{-i (\omega t-q_z Z)}
\nonumber \\
&&
	\times \bra{u_s} z_c  \ket{u_\pm} \delta_{s's}\delta_{n'n}
	\,,
\end{eqnarray}
where both LHs are equally excited. 
Note that for a QD smaller than the beam waist $E_z$ depends neither on $r$ nor on $\varphi$. Moreover, since the QD is assumed quasi 2D, to a good approximation the $\exp(i q_z Z)$ can be pulled out of the matrix element. The consequence is that no new selection rules exist, as signalled by the delta functions in the envelope wave-function quantum numbers. According to Eq.\ (\ref{Eq:LH_states}) $\bra{u_s} z_c  \ket{u_\pm} = 2/\sqrt{6} z_{cv}$ with $z_{cv}= \bra{s} z_c  \ket{p_z} \simeq 1$nm\cite{sergey2013optical, Bastard1988Wave}.
Since the second term in the LH decomposition has entered the calculation and the spin is not modified by the light-matter interaction, we see that the excitation produces a hole $\ket{3/2,+1/2}$ ($\ket{3/2,-1/2}$) with a conduction-band electron with spin down (up)--we remind that holes have opposite quantum numbers to electrons. Then, the excitation obviously produces pairs of $\text{AM}=0$. This is schematically represented in Fig. \ref{fig:Trans} by a dotted (red) line.

\subsection{Electron--light-hole pairs with angular momentum}
\label{Sec:OAM}

In Sect.\ \ref{Sec:AM0} we have seen that the $z$ component of the EM field of a TL beam having $\{\ell=1, \sigma=-1\}$ does not vanish at the phase singularity. This is a desirable feature, for it ensures a large fluency at the QD, and short exposure times would result in full transitions.
However, we notice that this set of parameters renders a $z$ component of the field with no angular dependence [see Eq.\ (\ref{Eq:AM0_Efield_z})]. 
The features of non-vanishing amplitude at $r=0$ and lack of angular dependence coexist in another set of values $\{\ell=2, \sigma=-1\}$ as one observes by inspection of the EM-fields expressions deduced in other works\cite{monteiro2009ang, klimov2012mapping, iketaki2007inv, quinteiro2014formulation}. 
It it unfortunate that fields having a finite amplitude at $r=0$ and would yield large fluency would not transfer {\it orbital} angular momentum to the material system.
In contrast, the in-plane components of Eq.\ (\ref{Eq:AM0_Efield_z}) and Eq.\ (\ref{Eq:Fields_OAM_cart}) with vanishing amplitude can indeed transfer orbital AM [in the former case, to account for the transition, one would need to add an interaction term to Eq.\ (\ref{Eq:H_AM0})].

It is our aim in this section to show that TL can be used to address the remaining eh pair states. By this we mean that eh pairs having band+spin $\text{AM}=1$ can be created. Moreover, by the use of $\ell\neq 0$ eh pairs with envelope AM are produced. 
One could consider the anti-parallel AM fields of Sect.\ \ref{Sec:AM0}, but the excitation is dominated by the $z$ component. We thus turn to the study of a beam having $\{\ell=1,\sigma=1\}$, in which case the $z$ component can be disregarded--other possible values of $\ell$ could be considered, and would produce transitions to other $n$ states, but the amplitude close to the origin of coordinates diminishes with increasing orbital AM\cite{quinteiro_79_155450}.

Following the discussion in Sect.\ \ref{Sec:AM0} we calculate the matrix elements corresponding to the absorption of a photon in the Rotating Wave Approximation from the interaction Hamiltonian $(-1/2) \mathbf d_\perp \cdot \mathbf E$ [Eq.\ (\ref{Eq:Ham_OAM})]
\begin{eqnarray}
\label{Eq:matrixE_OAM}
    \bra{\psi_{cs'n'}} H_I \ket{\psi_{LHsn}}
&=&
    -i \frac{q E_0 x_{vc}}{4\sqrt{6}} 
	e^{-i (\omega t-q_z Z)} 
\nonumber \\
&&
	\times
	\bra{s'n'} (q_r R) e^{i \varphi} \ket{sn} \delta_{LH,-}
\,,
\end{eqnarray}
where we used the separation of the matrix element into Bloch-periodic and envelope part, and the decomposition of LHs in $p$-like orbitals Eq.\ (\ref{Eq:LH_states}).
As in Eq.\ (\ref{Eq:AM0_ME}) $x_{vc}\simeq 1$nm. Since we have considered only the case where the polarization of the field is $\sigma = 1$, there is actually one possible transition, the one originating from $\ket{3/2, -1/2}$ as signalled by the delta function.
Using Eq.\ (\ref{Eq:complete_states}) the envelope-function matrix element is 
 \begin{eqnarray}
\label{Eq:matrixE_OAM_env}
    \bra{s'n'} (q_r R) e^{i \varphi} \ket{sn}
&=&
    \delta_{n+1,n'}
	\int_0^{\infty} dR \, (q_r R^2) 
\nonumber \\
&&
	\times
	{\cal R}_{is'n+1}^*(R) {\cal R}_{isn}(R)
\,.
\end{eqnarray}
Of main interest is the transition from $\{s=0,n\}$ to $\{s=0,(n+1)\}$. For a parabolic confinement potential with radial Laguerre-Gaussian wave-functions, the integral has analytical solution $(\sqrt{2}\pi)^{-1} q_r \ell_c \sqrt{n+1}$.
The product $q_r \ell_c$ makes explicit the importance of a well-focused beam. Twisted light can be effectively focused using an objective, so reducing the beam waist such that $q_r\simeq q_z$\cite{monteiro2009ang}. In this case the transition with $n=0$ (uppermost valence-band state) yields $\bra{01} (q_r R) e^{i \varphi} \ket{00}\simeq 3\times 10^{-2}$. 
As already discussed, the vanishing of the field at $r=0$ reduces the probability amplitude\cite{quinteiro_79_155450} making it necessary the use of longer and/or more intense pulses to complete a transition.

A different strategy to reduce the exposure time and intensity of the field is to use semiconductor systems with larger $E_g$. For example, QDs based on GaN/AlN are excited at $q_z\simeq300$nm$^{-1}$, so improving the factor $q_r \ell_c$. They confine both electrons and holes and can be modelled by Eq.\ (\ref{Eq:complete_states}) with Bessel radial wave-functions.

\section{Spintronic applications}
\label{Sec:spintronics}

In this section we will show how a single-beam configuration set-up using one
or multiple pulses of an optical-vortex field at normal incidence can be used to improve previous proposals using LH exciton dynamics in QDs, or can be used in new ways.

\subsection{Controlling magnetic impurities}
\label{Sec:}

In the past years several groups have been successful in placing a single magnetic ion, such as Mn or Co, in a single QD of either II-VI or III-V material\cite{besombes2004probing, kudelski2007optically, kobak2014designing}. The spin of such a magnetic impurity interacts with the electrons and holes in the QD; this interaction is usually well described by Heisenberg-type couplings between the various spins involved\cite{fernandez2006single, van2008single}. This interaction gives rise to a characteristic splitting of the exciton line of the QD into a multiplet of lines, the details being strongly dependent on both the QD material and the impurity atom\cite{besombes2004probing, kudelski2007optically, kobak2014designing, fernandez2006single, van2008single}.

The spin degree of freedom  of a Mn impurity in a II-VI QD has has been
proposed for applications in the field of spintronics or quantum information
processing due to its long lifetime\cite{le2010optical}. For this purpose it
is necessary to selectively prepare the Mn spin in any of its six possible
orientations. Various schemes have been proposed to achieve this task by
coherent manipulation of excitons
\cite{reiter2009all,reiter2011coherent,reiter2012spin}. These schemes are based on
combined spin-flips of electron and Mn spin or hole and Mn spin. In the case
of HH excitons, only electron spin flips are possible because a
Heisenberg-type Hamiltonian cannot produce transitions from a state with
$m_j=+3/2$ to $m_j=-3/2$ or vice versa. In contrast, in the case of LH
excitons both types of spin flips are possible, which makes this type of
excitons particularly attractive to achieve the Mn spin preparation.

The basic idea behind the Mn spin switching schemes is the following: By
optical excitation with a given polarization an exciton with a certain AM is
created. Then, a correlated dynamics of the Mn and the exciton spin system
starts and at a suitable later time when the exciton is in a state with a
different AM, this exciton is removed again by a light pulse with the
respective polarization. Restricting oneself to circularly polarized
plane-wave beams, only Mn spin flips by multiples of two are possible,
because only excitons with AM $\pm 1$ can be created or removed. However,
adding linearly polarized light with polarization along the $z$-direction
also LH excitons with AM $0$ can be created or removed and thus
the Mn spin can be driven into all of its six
eigenstates\cite{reiter2011coherent}. Using only plane-wave light, this scheme
however requires the application of light pulses propagating in normal
incidence direction and in an in-plane direction. From a practical point of
view, such a double-beam set-up is cumbersome, because it requires light
propagating in the QD plane and therefore typically cleaving of the sample is
needed. We propose here the use of a single normal-incidence beam having
either the values $\{\ell=0,\sigma=\pm 1\}$ to address LH excitons with AM
$\pm 1$, or $\{\ell=\pm 1,\sigma=\mp 1\}$ to address excitons
with AM 0. Switching between one and the other value of light's orbital AM
can be achieved by the use of two co-linear beams (a plane-wave one and the
other of TL) or by switching the topological charge of a single
beam\cite{mirhosseini2013rapid}.

In addition, twisted light offers a way to control the relative strength of
the interaction between carriers and impurities. This interaction is
typically well described by a contact interaction and its strength is
therefore proportional to the modulus square of the wave-function of the
involved carrier at the impurity position\cite{trojnar2012theo}. Electrons
and holes are usually excited by plane-waves, producing electron-hole pairs
where both carriers have the same orbital AM, e.g., $n=0$ in the $s$-shell or
$n=1$ in the $p$-shell. Thus, both carriers have their largest spatial
probability either close to the QD center ($s$-shell states) or away from the
center ($p$-shell states). Regardless of the impurity position--which varies
from sample to sample due to fabrication uncontrollable parameters--the ratio
between the electron-impurity and the hole-impurity interaction will be
essentially the same in all samples.

Consider instead the excitation by TL with $\{\ell=1,\sigma=+1\}$ as depicted
by dashed (blue) line in Fig.\ \ref{fig:Trans}. The electron and hole are
created in the envelope $p$ ($n=1$) and $s$ ($n=0$) states, respectively.
Likewise, by tuning the light frequency to the appropriate transition,
an electron-hole pair with electron in the envelope $s$ ($n=0$) and hole in
the envelope $p$ ($n=1$) state can be created. Thus, for an impurity placed
at the center, in the former case the interaction with the hole will be strongly
dominant while in the latter case the electron-impurity interaction will
dominate. Conversely, if the impurity is at some distance from the QD center,
in the first case the electron and in the second case the hole will interact
stronger. Magnetic fields can be used to further tune the transition frequency and therefore select which carrier (electron or hole) is excited into an orbital state with
$n\neq 0$\cite{quinteiro_79_155450}.

One can even imagine the use of superposition of twisted light beams having different $\ell$ to further narrow the localization region of carriers. 
Consider for example the simultaneous excitation with an $\{\ell=1, \sigma=1\}$ and an $(\ell=-1, \sigma=1)$ beams. If the laser frequency is tuned to the transition between the valence-band state $n=0$ and the conduction-band state $n=1$, the in-plane components of both fields will be responsible for the interaction. Then, an electron in a superposition state of envelope $n=+1$ and $n=-1$ with spin up will be generated. By introducing a relative phase between the two beams, the spatial orientation of the electron's wave-function can be rotated, thus allowing to better control the interaction strengths with an impurity located away from the QD center.


\subsection{Controlling the spin of an excess electron}
\label{Sec:}

Let us now consider the control of the spin of an excess  electron in a
charged QD by employing the $z$ component of the TL field. Pazy et
al.\cite{pazy2003spin} have suggested the use of a combination of two pulses
traveling in orthogonal directions to achieve such a spin flip. As was
previously mentioned, this idea faces implementation problems. For example, a
realistic quantum computing proposal based on semiconductor technology would
use a 2D array of a large number of charged QDs \cite{de2013ultrafast}, where
each excess electron realizes a qubit, and the spin-flip implements a Pauli-X
gate. For this proposal, the use of light beams propagating in the plane of
the array seems difficult or even impossible. However, as in other
situations\cite{sbierski2013twi}, the $z$-component of the optical vortex can
prove very handy by producing the spin-flip at normal incidence.

\begin{figure}[h]
  \centerline{\includegraphics[scale=.6]{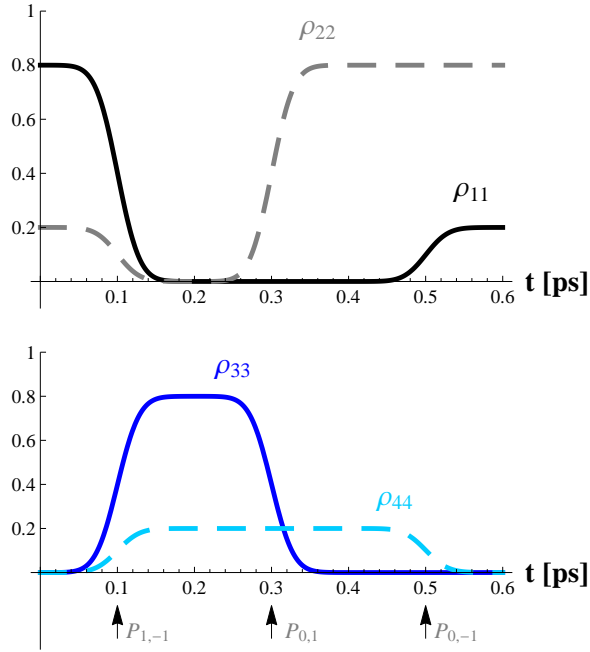}}
  \caption{Evolution of the density matrix from an initial pure state with $a_1^2=0.8$ and $a_2^2=0.2$. The sequence of ultrashort $\pi$-pulses $\{P_{1,-1}, P_{0,1}, P_{0,-1}\}$ is applied at times $t=0.1,0.3,0.5$ ps respectively; each gaussian pulse lasts for $\tau_p=50$ps.}
  \label{fig:evolution_v1}
\end{figure}

Relying on recent experiments\cite{yamane2012ultrashort}, we propose the use
of optical-vortex pulses $P_{\ell,\sigma}$ in the tens of femtoseconds range.
Note that the pulse duration is mainly limited by the spectral selectivity
required for a specific sample. A sequence of normal-incidence $\pi$-pulses:
$\{P_{1,-1} \rightarrow P_{0,1} \rightarrow P_{0,-1}\}$ (first pulse left)
excites the charged QD at normal-incidence.
\begin{figure}[h]
  \centerline{\includegraphics[scale=.7]{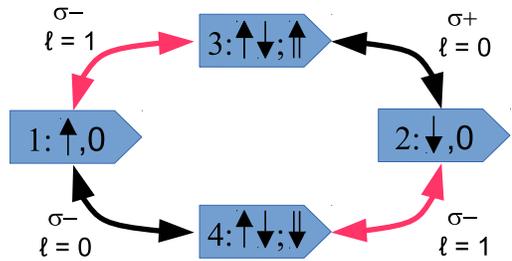}}
  \caption{Control of excess electron in charged QD, states and possible transitions. Arrows indicate electrons in conduction band; signs indicate light-hole AM projection.}
  \label{fig:states_01}
\end{figure}
The relevant states are the single electron states
$\{\ket{1}=\ket{\uparrow;0}, \ket{2}=\ket{\downarrow;0}\}$ and the charged
exciton (trion) states $\{\ket{3}=\ket{\uparrow\downarrow;\Uparrow},
\ket{4}=\ket{\uparrow\downarrow;\Downarrow}\}$, where the arrows
$\uparrow,\downarrow$ refer to the electron spin and $\Uparrow,\Downarrow$ to
the LH AM. Figure \ref{fig:states_01} shows a graphical representation of the
states and possible transitions. Such a pulse sequence can easily be seen to
produce spin inversion of the excess electron. For ultra-short excitation,
the whole evolution of the system can be broken into intervals of free
evolution and pulse excitation\cite{viola1998dynamical}. We will denote by
$U_{\ell,\sigma}$ the evolution operator for a certain time interval in which
an optical-vortex pulse $P_{\ell,\sigma}$ with pulse area $\pi$ is present.
Let us assume that the initial state of the system is $\ket{1}$. Applying to
it the aforementioned sequence of pulses, we get the final state
\begin{eqnarray}
\ket{\psi_f}&=&U_{0,-1}U_{0,1}U_{1,-1}\ket{1}
\nonumber \\
&=&U_{0,-1}U_{0,1}\ket{3}=U_{0,-1}\ket{2}=\ket{2},
\end{eqnarray}
disregarding the global phase arising from free evolution. On the other hand,
starting with the initial state $\ket{2}$, we get
\begin{eqnarray}
\ket{\psi_f}&=&U_{0,-1}U_{0,1}U_{1,-1}\ket{2} 
\nonumber \\
&=&U_{0,-1}U_{0,1}\ket{4}=U_{0,-1}\ket{4}=\ket{1}.
\end{eqnarray}
Thus, in both cases we obtain an inversion of the electron spin. We note
that, if the initial state is known, two pulses are sufficient, as in the
first case the third pulse has no action on the state and in the second case
this holds for the second pulse. The three-pulse sequence, however, works in
both cases and it also works for an arbitrary superposition of the two
electron spin states, as we see by a numerically simulation of the density
matrix $\rho(t)$ using the master-equation formalism within the 4-level
system. Figure \ref{fig:evolution_v1} shows the evolution starting from a
pure superposition state $\ket{\Psi}\bra{\Psi}$ with $\ket{\Psi} = a_1
\ket{1} + a_2 \ket{2}$. A full spin-inversion, equivalent to a Pauli-X gate,
is completed in less than a picosecond using moderate laser powers
($E_0\simeq10^9$V/m).
It is worth noting that there was no need to include decay and dephasing
processes. While the excitation and de-excitation of eh pairs takes place in
our protocol in less than a ps, the eh recombination and dephasing times are
in the hundreds of ps \cite{melliti2003radiative, mu2007dynamics}. In
addition, the spin dephasing time of conduction-band electron is in the nano-
to microsecond range \cite{feng2004spin, greilich2009optical, de2013ultrafast}.
However, some other factors are detrimental to the fidelity of the operation,
for instance the detuning of the laser and deviations of the pulse area from
the correct value $\pi$. This sensitivity to pulse parameters might be
overcome by employing the technique of adiabatic rapid passage using chirped
pulses\cite{wu2011population,simon2011robust,luker2012influence}.

Due to the short operation time, one could even relax the requirement of
energy inversion between light and heavy hole bands, and so the present
proposal would be compatible with QDs where the HH is the lowest hole state.
Finally, thanks to the fact that the $z$-component of the field is constant
near $r=0$ [the correction is of order $(q_r r)^2$], there is no need to
precisely center the beam axis on the QD.

\section{Conclusions}
\label{Sec:conc}

We have theoretically studied the excitation of light-holes in small quantum dots by highly-focused optical-vortex beams. To this end we employed the Twisted-light Gauge which casts the interaction Hamiltonian in terms of electric fields, and can account for the simultaneous action of all three spatial components that typically appear in tightly-focused beams.

We showed that a single optical-vortex beam at normal incidence can create all possible microscopic states of electron-hole pairs in the quantum dot, and we gave explicit expressions for the corresponding transition matrix elements. 

We applied our results to present proposals in spintronics, namely the manipulation of the spin of a magnetic impurity or of an excess electron in quantum dots, and demonstrated that those proposals can be improved by using optical-vortex fields. We also suggested a new possible way to address states of magnetic impurities, based on the transfer of orbital angular momentum to the envelope function of carriers in the quantum dot.

\begin{acknowledgments}
G.\ F.\ Quinteiro thanks the Argentine research agency CONICET for financial support through the ``Programa de Becas Externas''.
\end{acknowledgments}

\appendix*

\section{Neglecting magnetic terms}
\label{App:MagneticTerms}

In Ref.\ \onlinecite{quinteiro2014formulation} we have shown that a gauge transformation of the type
\begin{eqnarray}\label{Eq:chiGeneral}
	\chi (\mathbf r, t)
&=&
	- \frac{1}{|\ell| +1}  \mathbf r_\perp \cdot
	\mathbf A(\mathbf r,t)
\nonumber \\
&&
	- \frac{1}{|\ell+\sigma| +1}  z	A_z(\mathbf r,t)
\,,
\end{eqnarray}
leads to a convenient representation of the interaction Hamiltonian. For light having sign($\sigma$)$=$sign($\ell$), the Hamiltonian is given solely in terms of electric fields, while for sign($\sigma$)$\neq$sign($\ell$), one should in principle consider possible magnetic contributions, that are written in terms of the vector potential. 

We will next show that, for $\{\ell=1,\sigma=-1\}$ with the chosen parameters for QD and beam, the magnetic terms are in fact negligible. The gauge transformation applied to Eq.\ (\ref{Eq:A_Bessel}), produces a new scalar potential 
\begin{eqnarray}\label{Eq:UPGeneral}
	U'(\mathbf r, t)
&=&
	- \frac{1}{2}  \mathbf r_\perp \cdot \mathbf
	E(\mathbf r,t)
	- z	E_z(\mathbf r,t)
\,,
\end{eqnarray}
and a new vector potential
\begin{eqnarray}\label{Eq:A_TLgauge}
	A'_\varphi
&=&
	\frac{A_0}{2} (q_r r) \sin(\Phi) \,,
\nonumber \\
	A'_z
&=&
	 A_0 \frac{q_r}{q_z} (q_z z) \cos(\Phi)
	-\frac{A_0}{4} \frac{q_z}{q_r} (q_r r)^2 \sin(\Phi)
\,,
\end{eqnarray}
with $\Phi=\omega t - q_z z$.
In Sect.\ \ref{Sec:ell1} we have seen that the dominant component of the electric field is $E_z$ [see Eq.\ (\ref{Eq:AM0_Efield_z})]. Therefore, we are interested in comparing the magnetic interactions corresponding to Eq.\ (\ref{Eq:A_TLgauge}) with the $z$-component part of the electric interaction.

As an example, we provide details on the relative strength of magnetic and electric interactions arising from the $z$ components. We calculate the ratio of interband matrix element $\langle q (p_z/m)  A'_z\rangle$ to $\langle q z E_z\rangle$. In the former $A'_z$ acts on the envelope part of the wave-function, while $p_z$
acts on the Bloch-periodic part; thus, one can split the matrix element into $\bra{u_c} q (p_z/m) \ket{u_v} \bra{\phi_f{\cal Z}_f} A'_z\ket{\phi_i{\cal Z}_i}$. Furthermore, one can apply the well-known relationship $\bra{u_c} p_z \ket{u_v}= -i m \omega_{cv} \bra{u_c} z \ket{u_v}$. On the other hand, $E_z$ can be pulled out of its matrix element, since it does not depend on coordinates. We note that the magnetic and electic terms may connect different initial and final states, and they should all be compared. 
Let us next consider each term from $A'_z$ separately:
\begin{eqnarray}\label{Eq:ratios_1}
	\frac{\langle q (p_z/m)  A'_{z1}\rangle}{\langle q z E_z\rangle} 
&=&
	\omega_{cv} \frac{\bra{\phi_f{\cal Z}_f} A'_{z1}
	\ket{\phi_i{\cal Z}_i}}{E_z}
\nonumber \\
&=&
	\frac{q_r}{q_z} \frac{\omega_{cv}}{\omega}
	\bra{\phi_i{\cal Z}_f}(q_z z)\ket{\phi_i{\cal Z}_i}
\!.
\end{eqnarray}
The operator $z$ forces initial and final states to be of opposite parity. Since the QD is quasi 2D, the next $z$-subband has a much larger energy, and thus is well detuned from the transitions considered in the present paper. Moreover, the matrix element is proportional to $(q_z z)\ll 1$. Thus, it is safe to neglect the first term of $A'_z$. 
Next, we look at the second term
\begin{eqnarray}\label{Eq:ratios_1}
	\frac{\langle q (p_z/m)  A'_{z2}\rangle}{\langle q z E_z\rangle} 
&=&
	\omega_{cv} \frac{\bra{\phi_f{\cal Z}_f} A'_{z2}
	\ket{\phi_i{\cal Z}_i}}{E_z}
\nonumber \\
&=&
	\frac{1}{4} \frac{q_z}{q_r} \frac{\omega_{cv}}{\omega}
	\bra{\phi_f{\cal Z}_i}(q_r r)^2\ket{\phi_i{\cal Z}_i}
\!.
\end{eqnarray}
Since the $\ket{\phi{\cal Z}}$ are not eigenstate of the operator $r$, we can only argue that the ratio is proportional to $(q_r r)^2\ll 1$, and thus is also negligible. In conclusion, we see that the magentic field interaction is much smaller than the electric interaction considered in the article.

Similar arguments can be given for $(q/m) p_\varphi A_\varphi$ and $(-1/2) \mathbf d_\perp \cdot \mathbf E$. As a result, the Hamiltonian reads
\begin{eqnarray}\label{Eq:H_APapprox}
	H
&=&
	\frac{1}{2m} \mathbf p^2 + V(\mathbf r)
    - d_z E_z(\mathbf r,t)
\,.
\end{eqnarray}
%

%

\end{document}